\documentclass[9pt,technote]{IEEEtran}

\usepackage{cite}      

\usepackage{graphicx}  

\usepackage{url}       

\usepackage{amsmath}   
\newtheorem{theorem}{\textbf{Theorem}}
\newtheorem{definition}{\textbf{Definition}}
\newtheorem{corollary}[theorem]{\textbf{Corollary}}
\newtheorem{example}{\textbf{Example}}
\newtheorem{lemma}[theorem]{\textbf{Lemma}}
\newtheorem{algorithm}{\textbf{Algorithm}}

\begin{document}
%
\title{Steganographic Codes --- a New Problem of \newline Coding Theory }
%
%
\author{Weiming~Zhang, and~Shiqu~Li 
\thanks{Weiming Zhang and Shiqu Li are both with the Department of Applied Mathematics,
University of Information Engineering, P.O. Box 1001-747,
Zhengzhou 450002 P.R. China. Email: nlxd\_990@yahoo.com.cn}}
%
%
%
\markboth{Journal of \LaTeX\ Class Files,~Vol.~1, No.~11,~November~2002}{Shell \MakeLowercase{\textit{et al.}}: Bare Demo of IEEEtran.cls for Journals}
%



\maketitle

\begin{abstract}
To study how to design steganographic algorithm more efficiently,
a new coding problem -- steganographic codes (abbreviated
stego-codes) -- is presented in this paper. The stego-codes are
defined over the field with $q(q\ge2)$ elements. Firstly a method
of constructing linear stego-codes is proposed by using the direct
sum of vector subspaces. And then the problem of linear
stego-codes is converted to an algebraic problem by introducing
the concept of $t$th dimension of vector space. And some bounds on
the length of stego-codes are obtained, from which the maximum
length embeddable (MLE) code is brought up. It is shown that there
is a corresponding relation between MLE codes and perfect
error-correcting codes. Furthermore the classification of all MLE
codes and a lower bound on the number of binary MLE codes are
obtained based on the corresponding results on perfect codes.
Finally hiding redundancy is defined to value the performance of
stego-codes.
\end{abstract}


%
\IEEEpeerreviewmaketitle

\section{Introduction}
%
%
%
%
\PARstart{N}owadays the security of communication means not only
secrecy but also concealment, so steganography is becoming more
and more popular in the network communication. Steganography is
about how to send secret message covertly by embedding it into
some innocuous cover-objects such as digital images, audios and
videos. In this paper we take the image as example to describe our
ideas. Usually the process of embedding message will make some
changes to the cover-images. To reduce the possibility of
detection, the sender hopes to embed as most bits of message as
possible by changing the least number of bits of images. This task
can be accomplished through some encoding technique that is
firstly brought up by Crandall \cite{cra} who call it matrix
encoding. And in the present paper we generalize the idea of
Crandall and formally define this kind of codes as
``steganographic codes'' (abbreviated stego-codes).

Besides increasing the embedding efficiency, stego-codes can also
enhance the security of steganography at other aspects. Now some
detecting methods on steganography can not only detect the
existence of the hidden message but also very accurately estimate
its length \cite{zha:pin,jf:mg}. And there is even methods which
can search for the stego-key \cite{fri:gol:sou}. However, if there
are a great many stego-codes that can be selected by the encoders
as a part of the key, it will be very hard for the attacker to
estimate the message length or recovery the stego-key. In fact,
Fridrich \cite{fri:gol:sou} ever pointed out that matrix encoding
is an effective measure against key search.

LSB (Least Significant Bit) steganography is the most popular
image steganographic technique. by simple LSB steganography the
encoder selects a pixel (or DCT coefficient) every time and embeds
one bit of message in its LSB by modifying methods such as
replacing or $\pm 1$. This traditional technique can be viewed as
coding two bits of message per changed pixel because in random
case 50\% pixels needn't to be changed. A better method is
described in the CPT scheme \cite{che:pan,Tse:pan}, which is a
steganographic algorithm on binary image and can conceal as many
as $k$ bits of data in a host image of size $2^k-1$ by changing at
most 2 bits. Another more effective example of stego-code is F5
\cite{wes}, a LSB algorithm on JPEG image, which firstly
implements Crandall's matrix encoding and can embed $k$ bits of
message in $2^k-1$ DCT coefficients by changing at most one of
them.

To construct more effective stego-codes and study their
properties, in the present paper we define linear stego-codes over
finite field with $q(q\ge2)$ elements by using multi-outputs logic
functions. Firstly, as an example, a constructive method of linear
stego-codes is proposed, which can generate the codes of F5 in a
special case and is more agile than the codes of CPT. To study
bounds on the length of linear stego-codes, we introduce the
definition of $t$th dimension of vector space that converts the
problems of linear stego-codes to pure algebraic problems. And
then a bound on the length of linear stego-codes is obtained, from
which we bring out the maximum length embeddable (abbreviated MLE)
codes. Furthermore, it is shown that there is a 1-1 correspondence
between linear MLE codes and linear perfect error-correcting
codes.

To study the nonlinear stego-code, another direct definition for
stego-codes is presented, based on which we explain the relations
and differences between stego-codes and error-correcting codes in
geometrical language and generalize linear MLE codes to nonlinear
case. We prove the relations between MLE codes and perfect codes
with two constructive proofs which can be used to construct MLE
codes by perfect codes or construct perfect codes by MLE codes.
Furthermore from the well-known results on perfect codes, the
classification of all MLE codes and a lower bound on the number of
binary MLE codes are obtained.

Usually a steganographic algorithm can be valued by both message
rate and change density. Large message rate and small change
density means a good algorithm. To evaluate the performance of
stego-codes more accurately, we introduce the concept of hiding
redundancy that can be viewed as a combination of message rate and
change density. Furthermore based on the result on hiding
redundancy, another bound on the length of binary stego-codes is
obtained.

The rest of the paper is organized as follows. The construction
and properties of linear stego-codes are analyzed in Sect. II.
Nonlinear stego-codes and the relations between the MLE codes and
perfect codes are studied in Sect. III. In Sect. IV a measure --
hiding redundancy -- is proposed to value the efficiency of
stego-codes. And the paper concludes with a discussion in Sect. V.

\section{Linear Stego-codes}
\subsection{Definitions}

To deal with the concepts that are introduced we adopt some
notational conventions that are commonly used. The finite field
with $q$ elements is denoted by $GF(q)$. The vector is denoted by
bolt italic letter (e.g. $\textbf{\emph{x}}$). The set is denoted
by script letters (e.g. $\mathcal{S}$). And denote the Hamming
weight of a vector $\textbf{\emph{x}}\in GF^n(q)$ as
${\mbox{Wt}}({\textbf{\emph{x}}})$.

For simpleness, we take LSB steganography on images as examples to
describe the definitions and applications of stego-codes.

\medskip
\begin{definition}\label{def-1}
An $(n,k,t)$ stego-coding function over finite field $GF(q)$ is a
vectorial function $H({\textbf{\emph{x}}}) = (h_1
({\textbf{\emph{x}}}),h_2 ({\textbf{\emph{x}}}), \cdots ,h_k
({\textbf{\emph{x}}})): \, GF^n (q) \to GF^k (q)$ satisfying the
following condition: For any given $\textbf{\emph{x}} \in GF^n
(q)$ and ${\textbf{\emph{y}}}\in GF^k (q)$, there exists a
$\textbf{\emph{z}} \in GF^n (q)$ such that
$\rm{Wt}(\textbf{\emph{z}}) \le t$ and $H(\textbf{\emph{x}} +
\textbf{\emph{z}}) = \textbf{\emph{y}}$. And
$H(\textbf{\emph{x}})$ is called linear stego-coding function if
every component function $h_i(\textbf{\emph{x}})$ $(1 \le i \le
n)$ is a linear function.
\end{definition}
\medskip

\begin{definition}\label{def-2}
Let $H({\textbf{\emph{x}}})$ is an $(n,k,t)$ stego-coding function
over $GF^n (q)$. And for $\textbf{\emph{y}}\in GF^k (q)$, let
$H^{- 1} (\textbf{\emph{y}}) = \{
\textbf{\emph{x}}:\,H(\textbf{\emph{x}}) = \textbf{\emph{y}}\}$.
Then call
\[
\mathcal{S} = \{ H^{-1}(\textbf{\emph{y}}):\,\textbf{\emph{y}} \in
GF^k (q)\,\mbox{and}\,H^{- 1} (\textbf{\emph{y}}) \ne \phi \}
\]
an $(n,k,t)$ stego-code.
\end{definition}
\medskip

Stego-coding function in principle is the decoding function, and
to hide message with it, one also need an encoding algorithm.
Generally, encoding algorithm can be implemented through an
encoding table $\textbf{\emph{B}}$. For an $(n,k,t)$ stego-coding
function $H({\textbf{\emph{x}}})$ over $GF(q)$, encoding table
$\textbf{\emph{B}}$ is a $q^n \times q^k $ matrix, the index of
its row is represented by $\textbf{\emph{x}} \in GF^n (q)$, and
the index of a column by $\textbf{\emph{y}} \in GF^k (q)$. In the
position $(\textbf{\emph{x}},\textbf{\emph{y}})$, save the vector
$\textbf{\emph{z}} \in GF^n (q)$ such that
$\rm{Wt}(\textbf{\emph{z}}) \le t$ and $H(\textbf{\emph{x}} +
\textbf{\emph{z}}) = \textbf{\emph{y}}$. If
$H({\textbf{\emph{x}}})$ is a linear stego-coding function,
because $H(\textbf{\emph{x}} + \textbf{\emph{z}})
=H(\textbf{\emph{x}})+H(\textbf{\emph{z}})$, one only need
construct a $1 \times q^k $ encoding table, and denote the index
of a column with $\textbf{\emph{y}} \in GF^k (q)$. In position
$\textbf{\emph{y}}$, save the vector $\textbf{\emph{x}} \in GF^n
(q)$ such that $\rm{Wt}(\textbf{\emph{x}}) \le t$ and
$H(\textbf{\emph{x}})=\textbf{\emph{y}}$. Therefore for linear
stego-codes generally there exists simpler encoding algorithm.
Crandall points out that the design of fast encoding algorithm are
also an open research area \cite{cra}. The following example shows
a wonderful encoding method.

\medskip
\begin{example}[\rmfamily {F5-Matrix Coding}]\label{exa-1}
F5 \cite{wes} is a LSB steganographic program that embeds binary
message sequences into the LSBs of DCT coefficients of JPEG
images. F5 can embed $k$ bits of message in $2^k-1$ DCT
coefficients by changing at most one of them. The inputs are code
word (LSBs of DCT coefficients) $\textbf{\emph{x}}\in GF^{2^k  -
1}(2)$ and the block of message $\textbf{\emph{y}} \in GF^k (2)$.
The coding function is defined as
\begin{equation}\label{eqn-1}
f(\textbf{\emph{x}}) = \mathop  \oplus \limits_{i = 1}^{2^k  - 1}
x_i \cdot i \enspace ,
\end{equation}
where, to do $\oplus$, the integer $x_i \cdot i$ is interpreted as
a binary vector. And the encoding procedure is as follows: Compute
the bit place that has to be changed as $s = \textbf{\emph{y}}
\oplus f(\textbf{\emph{x}})$ where the resulting binary vector $s$
is interpreted as an integer. And then output the changed code
word
\[
\textbf{\emph{x}}'= \left\{ \begin{array}{ll}
\textbf{\emph{x}} & \quad \mbox{if } s = 0 \\
(x_1 ,x_2 , \cdots ,x_s \oplus 1, \cdots ,x_{2k + 1} ) &\quad
\mbox{if } s \ne 0
\end{array} \right.
\]
which satisfies $\textbf{\emph{y}}=f(\textbf{\emph{x}}')$.

According to Definition \ref{def-1}, (\ref{eqn-1}) in fact is a
$(2^k-1,k,1)$ linear stego-coding function over $GF(2)$. For
instance, when $k=2$, (\ref{eqn-1}) is equivalent to the vectorial
function $H(\textbf{\emph{x}}) = (h_1
(\textbf{\emph{x}}),\,h_2(\textbf{\emph{x}}))$ where ($h_1
(\textbf{\emph{x}}) = x_2 \oplus x_3 $, $h_2(\textbf{\emph{x}}) =
x_1 \oplus x_3$). And the corresponding stego-code is
\begin{eqnarray*}
\mathcal{S} = &\{&\{(000),(111)\},\,\{(011),(100)\} ,\\
&&\{(010),(101)\} ,\,\{(001),(110)\} \enspace\,\}\enspace .
\end{eqnarray*}
\end{example}
\bigskip

CPT scheme \cite{che:pan,Tse:pan} is an example of nonlinear $(2^k
- 1,k,2)$ stego-coding function. We firstly study linear
stego-coding function which has the following necessary and
sufficient condition.

\medskip
\begin{theorem}\label{thm-1}
Linear vectorial function $H(\textbf{\emph{x}})$ over $GF(q)$ is
an $(n,k,t)$ stego-coding function if and only if for any given
$\textbf{\emph{y}}\in GF^k(q)$, there exists a $\textbf{\emph{z}}
\in GF^n (q)$ such that $\rm{Wt}(\textbf{\emph{z}})\le t$ and
$H(\textbf{\emph{z}})=\textbf{\emph{y}}$.

\begin{proof}
If $H(\textbf{\emph{x}})$ is a linear stego-coding function over
$GF(q)$, Definition \ref{def-1} implies that for any given
$\textbf{\emph{y}}\in GF^k(q)$ and $\textbf{\emph{0}}\in GF^k(q)$,
there exists a $\textbf{\emph{z}} \in GF^n (q)$ such that
$\rm{Wt}(\textbf{\emph{z}})\le t$ and
$\textbf{\emph{y}}=H(\textbf{\emph{0}}+\textbf{\emph{z}})=H(\textbf{\emph{z}})$.

Conversely, for any given $\textbf{\emph{x}}\in GF^n(q)$ and
$\textbf{\emph{y}}\in GF^k(q)$ there exists a $\textbf{\emph{z}}
\in GF^n (q)$ such that $\rm{Wt}(\textbf{\emph{z}})\le t$ and
$H(\textbf{\emph{z}}) = \textbf{\emph{y}} - H(\textbf{\emph{x}})$,
i.e. $H(\textbf{\emph{x}} + \textbf{\emph{z}}) =
\textbf{\emph{y}}$ because $H(\textbf{\emph{x}})$ is a linear
function. Therefore $H(\textbf{\emph{x}})$ satisfies the condition
of Definition \ref{def-1}.
\end{proof}
\end{theorem}
\medskip

An $(n,k,t)$ linear vectorial function $H({\textbf{\emph{x}}}) =
(h_1 ({\textbf{\emph{x}}}), h_2 ({\textbf{\emph{x}}}), \cdots ,
h_k ({\textbf{\emph{x}}}))$ over $GF(q)$, where
$h_i(\textbf{\emph{x}}) = a_{i1} x_1  + a_{i2} x_2  + \cdots +
a_{in} x_n $ $(1 \le i \le k)$ can be represented by a $k \times
n$ matrix over $GF(q)$ such as
\[
\textbf{\emph{H}}=\left[ \begin{array}{cccc}
a_{11} & a_{12} & \cdots & a_{1n}\\
a_{21} & a_{22} & \cdots & a_{2n}\\
\cdots \\
a_{k1} & a_{k2} & \cdots & a_{kn}\end{array} \right]\enspace .
\]
We call $\textbf{\emph{H}}$ an $(n,k,t)$ stego-coding matrix.
There is a 1-1 correspondence between stego-coding functions and
stego-coding matrices. And from Theorem \ref{thm-1}, we can define
the stego-coding matrix directly as follows.

\medskip
\begin{definition}\label{def-3}
A $k \times n$ matrix $\textbf{\emph{H}}$ over $GF(q)$  is called
stego-coding matrix if for any given $\textbf{\emph{y}} \in GF^k
(q)$, there exists an $\textbf{\emph{x}} \in GF^n (q)$ such that
$\rm{Wt}(\textbf{\emph{x}}) \le t$ and
$\textbf{\emph{H}}\textbf{\emph{x}}^{\rm{tr}}= \textbf{\emph{y}}$.
\end{definition}
\medskip

If $\textbf{\emph{H}}$ is an $(n,k,t)$ stego-coding matrix over
$GF(q)$, then for any $\textbf{\emph{y}} \in GF^k (q)$, equation
$\textbf{\emph{H}}\textbf{\emph{x}}^{\rm{tr}}= \textbf{\emph{y}}$
has solutions, which implies that the rank of $\textbf{\emph{H}}$
is $k$. From Definition \ref{def-3} we can get the following
important property that is useful for the construction of linear
stego-coding functions.

\medskip
\begin{theorem}\label{thm-2}
A $k \times n$ matrix $\textbf{\emph{H}}$ over $GF(q)$  is an
$(n,k,t)$ stego-coding matrix if and only if, for any
$\textbf{\emph{y}} \in GF^k (q)$, $\textbf{\emph{y}}^{\rm{tr}}$
must be a linear combination of some $t$ columns of
$\textbf{\emph{H}}$.
\end{theorem}

\subsection{A Constructing Method of Linear Stego-coding Functions}

Theorem \ref{thm-2} suggests that we can construct stego-coding
matrix through the direct sum of vector subspaces. To do that, we
need the following lemma.

\medskip
\begin{lemma}\label{lem-3}
If $V$ is a $k$-dimensional vector space over $GF(q)$ then there
exists $\scriptstyle\frac{q^k -1}{q - 1}$ vectors
$\textbf{\emph{x}}_1, \cdots,
\textbf{\emph{x}}_{\scriptstyle\frac{q^k -1}{q - 1}}$ satisfying
the following properties:
\begin{enumerate}
\item Any two of the $\scriptstyle\frac{q^k -1}{q - 1}$ vectors
are linear independence. \item For any given $\textbf{\emph{y}}
\in V$, there exist $a \in GF(q)$ and $\textbf{\emph{x}}_i$, such
that $1 \le i \le \scriptstyle\frac{q^k - 1}{q - 1}$ and
$\textbf{\emph{y}}=a \textbf{\emph{x}}_i$.
\end{enumerate}
\begin{proof}
Take any nonzero vector $\textbf{\emph{x}}_1 \in V$, and denote
the 1-dimensional subspace spanned by $\textbf{\emph{x}}_1$ as
$V_1$; then take any nonzero vector $\textbf{\emph{x}}_2  \in
V\backslash V_1 $ and denote the 1-dimensional subspace spanned by
$\textbf{\emph{x}}_2$ as $V_2$; and then take any nonzero vector
$\textbf{\emph{x}}_3 \in V\backslash (V_1 \cup V_2 )\cdots$. Do as
such and finally we can get $\scriptstyle\frac{q^k -1}{q - 1}$
1-dimensional subspaces $V_1, \cdots, V_{\scriptstyle\frac{q^k
-1}{q - 1}}$ because the number of nonzero vectors in $V$ is
$q^k-1$ and every 1-dimensional subspace consist of $q-1$ nonzero
vectors and the zero vector. Assume that subspace $V_i$ is spanned
by $\textbf{\emph{x}}_i$ $(1 \le i \le \scriptstyle\frac{q^k -
1}{q - 1}\displaystyle)$, The procedure of constructing these
subspaces implies that any two of these $\textbf{\emph{x}}_i$'s
are linear independence and $V=V_1 \cup V_2 \cup \cdots \cup
V_{\scriptstyle\frac{q^k -1}{q - 1}}$. Therefore for any given
$\textbf{\emph{y}} \in V$, there is $V_i$ satisfying
$\textbf{\emph{y}} \in V_i$, which means there exists $a \in
GF(q)$ such that $\textbf{\emph{y}}=a \textbf{\emph{x}}_i$.
\end{proof}
\end{lemma}
\medskip

Based on Lemma \ref{lem-3}, we can get the following constructive
algorithm of $(\sum_{i = 1}^t {\frac{q^{k_i} - 1}{q -
1}}\displaystyle,k,t)$ stego-coding matrix over $GF(q)$.

\medskip
\begin{algorithm}\label{alg-1}
The procedure of construction goes through the following three
steps.
\begin{description}
\item[S1] Take a basis of $k$-dimensional vector space $GF^k(q)$
over $GF(q)$ such as $\{\textbf{\emph{x}}_1, \textbf{\emph{x}}_2,
\cdots, \textbf{\emph{x}}_k \}$. \item[S2] Divide
$\{\textbf{\emph{x}}_1, \textbf{\emph{x}}_2,\cdots,
\textbf{\emph{x}}_k \}$ into $t$ disjoint subsets $B_i$ $(1\le
i\le t)$ such that $B_i$ consists of $k_i$ vectors and $\sum_{i =
1}^t {k_i }= k$. Denote the $k_i$ -dimensional subspace spanned by
$B_i$ as $V_i$, $1 \le i \le t$. \item[S3] As doing in the proof
of Lemma \ref{lem-3}, take $\scriptstyle\frac{q^{k_i} -1}{q - 1}$
nonzero vectors from every subspace $V_i$ $(1 \le i \le t)$. And
we can get $\scriptstyle\sum_{i = 1}^t {\frac{q^{k_i} - 1}{q -
1}}$ nonzero vectors in all. Then construct a $k \times
\scriptstyle\sum_{i = 1}^t {\frac{q^{k_i} - 1}{q - 1}}$ matrix
$\textbf{\emph{H}}$ with all of these nonzero vectors as columns.
And $\textbf{\emph{H}}$ is just a $(\scriptstyle\sum_{i = 1}^t
{\frac{q^{k_i} - 1}{q - 1}}\displaystyle,k,t)$ stego-coding matrix
over $GF(q)$.
\end{description}
\end{algorithm}
\medskip

In fact by Lemma \ref{lem-3}, for any subspace $V_i$ and any
vector $\textbf{\emph{x}} \in V_i $ in Algorithm \ref{alg-1},
there exists a column of $\textbf{\emph{H}}$ which can linearly
express $\textbf{\emph{x}} ^{\rm{tr}}$. On the other hand,
$GF^k(q)$ is the direct sum of these $t$ subspaces $V_i$'s.
Combine these two facts, it can be proved that, for any
$\textbf{\emph{y}}\in GF^k(q)$, $\textbf{\emph{y}}^{\rm{tr}}$ is
the linear combination of $t$ columns of $\textbf{\emph{H}}$.
Therefore by Theorem \ref{thm-2}, $\textbf{\emph{H}}$ is a
$(\scriptstyle\sum_{i = 1}^t {\frac{q^{k_i} - 1}{q -
1}}\displaystyle,k,t)$ stego-coding matrix over $GF(q)$.

Let $q=2$ and $t=1$, with Algorithm \ref{alg-1} we can construct
$(2^k-1,k,1)$ linear stego-coding functions over $GF(2)$ which are
just the functions used in F5 (Example \ref{exa-1}).

\subsection{The $t$th Dimension of Vector Space -- Bounds on the length of Linear
Stego-codes}

To study bounds on the length of stego-codes, we generalize the
concept of vector space's dimension to define the $t$th dimension.

\medskip
\begin{definition}\label{def-4}
If $V$  is a vector space over field $F$, $\textbf{\emph{x}},
\textbf{\emph{x}}_1, \textbf{\emph{x}}_2, \cdots,
\textbf{\emph{x}}_n \in V$ and there are $a_1, a_2, \cdots, a_n
\in F$ such that $\rm{Wt}\left((a_1 ,a_2 , \cdots ,a_n )\right)
\le t$ and $\textbf{\emph{x}} = \sum_{i=1}^n {a_i
\textbf{\emph{x}}_i }$, we say that $\textbf{\emph{x}}$ can be
expressed as $t$th linear combination of $\textbf{\emph{x}}_i$'s;
If  for any $\textbf{\emph{x}} \in V$, $\textbf{\emph{x}}$ can be
expressed as $t$th linear combination of $\textbf{\emph{x}}_i$'s,
we say that $\{\textbf{\emph{x}}_1, \textbf{\emph{x}}_2, \cdots,
\textbf{\emph{x}}_n\}$ is a set of $t$th generators of $V$.
\end{definition}
\medskip

\begin{definition}\label{def-5}
Let $V$ is a vector space over field $F$ and
$\{\textbf{\emph{x}}_1, \textbf{\emph{x}}_2, \cdots,
\textbf{\emph{x}}_n\}$ is a set of $t$th generators of $V$. If any
another set of $t$th generators $\{\textbf{\emph{y}}_1,
\textbf{\emph{y}}_2, \cdots, \textbf{\emph{y}}_m\}$ must satisfy
that $m \ge n$, we call $\{\textbf{\emph{x}}_1,
\textbf{\emph{x}}_2, \cdots, \textbf{\emph{x}}_n\}$ a minimum set
of $t$th generators of $V$ and call $n$ the $t$th dimension of
$V$.
\end{definition}
\medskip

In the terms of $t$th dimension, Theorem \ref{thm-2} can be stated
in the following forms.

\medskip
\begin{theorem}\label{thm-4}
A $k\times n$ matrix $\textbf{\emph{H}}$ is an $(n,k,t)$
stego-coding matrix over $GF(q)$ if and only if the set consisting
of $n$ vectors corresponding to the $n$  columns of
$\textbf{\emph{H}}$ is a set of $t$th generators of $GF^k(q)$.
\end{theorem}
\medskip

Because a set of $t$th generators must be a set of $(t+1)$th
generators, it is clear that for vector space $GF^k(q)$ and $t$
such that $t \ge k$, the $t$th dimension is $k$, and every basis
of $GF^k(q)$ is just a minimum set of $t$th generators of
$GF^k(q)$. In fact the $t$th dimension of $GF^k(q)$ such that $t >
k$ is insignificant for the problem of stego-codes.

The following theorem is easy to be get but is important, because
it converts the problem of linear stgeo-codes to a pure algebraic
problem.

\medskip
\begin{theorem}\label{thm-5}
If the $t$th dimension of vector space $GF^k(q)$ over $GF(q)$ is
$n$, then for any integer $m \ge n$ there exist $(m,k,t)$ linear
stego-codes.
\end{theorem}
\medskip

From Theorem \ref{thm-5}, we know that the key problems of linear
stego-codes are just how to estimate the $t$th dimension of
$GF^k(q)$ and how to construct the minimum set of $t$th generators
of $GF^k(q)$. Generally, it is hard to get the exact $t$th
dimension of $GF^k(q)$, but we can obtain some bounds on it, which
is also the bounds on the length of linear stego-codes.

\medskip
\begin{theorem}\label{thm-6}
If the $t$th dimension of vector space $GF^k(q)$ over $GF(q)$ is
$n$, then
\begin{equation}\label{eqn-2}
q^k  \le 1 + (q - 1){n \choose 1} + (q - 1)^2 {n \choose 2}   +
\cdots  + (q - 1)^t {n \choose t}\enspace .
\end{equation}
\begin{proof}
Assume that $\{\textbf{\emph{x}}_1, \textbf{\emph{x}}_2, \cdots,
\textbf{\emph{x}}_n\}$ is a set of $t$th generators of $GF^k(q)$.
Then for any $\textbf{\emph{x}} \in GF^k(q)$, $\textbf{\emph{x}}$
can be expressed as $t$th linear combination of
$\{\textbf{\emph{x}}_1, \textbf{\emph{x}}_2, \cdots,
\textbf{\emph{x}}_n\}$. On the other hand, there are in total $1 +
(q - 1){n \choose 1} + (q - 1)^2 {n \choose 2}   + \cdots  + (q -
1)^t {n \choose t}$ $t$th linear combinations of
$\{\textbf{\emph{x}}_1, \textbf{\emph{x}}_2, \cdots,
\textbf{\emph{x}}_n\}$ and $q^k$ vectors in $GF^k(q)$. Therefore,
we get the inequality (\ref{eqn-2}).
\end{proof}
\end{theorem}
\medskip

As mentioned above the $k$th dimension of vector space $GF^k(q)$
over $GF(q)$ is $k$, so when $t=k$ the equality holds in
(\ref{eqn-2}). And the following corollary shows that the equality
also holds in (\ref{eqn-2}) with $t=1$.

\medskip
\begin{corollary}\label{cor-7}
The first dimension of vector space $GF^k(q)$ over $GF(q)$ is
$\scriptstyle\frac{q^k -1}{q - 1}$,  and any set consisting of
$\scriptstyle\frac{q^k -1}{q - 1}$ vectors such that any two of
them are linear independence is a minimum set of the first
dimension generators.

\begin{proof}
For any given $\textbf{\emph{x}}_1, \cdots,
\textbf{\emph{x}}_{\scriptstyle\frac{q^k -1}{q - 1}} \in GF^k(q)$
such that any two of them are linear independence, the proof of
Lemma \ref{lem-3} means that $\big\{\textbf{\emph{x}}_1, \cdots,
\textbf{\emph{x}}_{\scriptstyle\frac{q^k -1}{q - 1}}\big\}$ is a
set of the first generators of $GF^k(q)$. Because when
$n=\scriptstyle\frac{q^k -1}{q - 1}$ and $t=1$, the equality in
(\ref{eqn-2}) holds, $\big\{\textbf{\emph{x}}_1, \cdots,
\textbf{\emph{x}}_{\scriptstyle\frac{q^k -1}{q - 1}}\big\}$ is a
minimum set of the first generators. Therefore the first dimension
of vector space $GF^k(q)$ is $\scriptstyle\frac{q^k -1}{q - 1}$.
\end{proof}
\end{corollary}
\medskip

By Lemma \ref{lem-3}, Corollary \ref{cor-7} and Theorem
\ref{thm-4}, for any $q \ge 2$ and $k \ge 1$, the
$(\scriptstyle\frac{q^k -1}{q - 1}\displaystyle,k,1)$ linear
stego-codes over $GF(q)$ exist, and when $q=2$, we get the codes
of F5 once more.

By Theorem \ref{thm-4} and \ref{thm-6}, an $(n,k,t)$ linear
stego-code over $GF(q)$ must satisfy (\ref{eqn-2}), which provides
a upper bound on the embedded message length. Therefore when
equality holding in (\ref{eqn-2}), we get an important type of
codes.

\medskip
\begin{definition}\label{def-6}
An $(n,k,t)$ linear stego-code over $GF(q)$ is called maximum
length embeddable (abbreviated MLE) if equality holds in
(\ref{eqn-2})
\end{definition}
\medskip

Note that the form of the bound in Theorem \ref{thm-6} is similar
with that of Hamming Bound on error-correcting codes.

\medskip
\begin{lemma}[Hamming Bound]\label{lem-8}
A $t$-error-correcting $(n,k)$ linear code over $GF(q)$ must
satisfy that
\begin{equation}\label{eqn-3}
q^{n-k} \ge 1 + (q - 1){n \choose 1} + (q - 1)^2 {n \choose 2}   +
\cdots + (q - 1)^t {n \choose t} \enspace .
\end{equation}
\end{lemma}
\bigskip

Error-correcting codes are called perfect codes when equality
holds in (\ref{eqn-3}). The Crandall's examples \cite{cra}, which
are obtained from perfect codes, are just linear MLE codes. The
following theorem will show the relations between linear MLE codes
and linear perfect codes.

\medskip
\begin{theorem}\label{thm-9}
An $(n-k)\times n$ matrix $\textbf{\emph{H}}$ is the parity check
matrix of a $t$-error-correcting perfect $(n,k)$ code over $GF(q)$
if and only if $\textbf{\emph{H}}$ is a stego-coding matrix of an
$(n,n-k,t)$ MLE code over $GF(q)$.

\begin{proof}
If $\textbf{\emph{H}}$ is the parity check matrix of a
t-error-correcting code, any two $t$th linear combinations of the
n columns of $\textbf{\emph{H}}$ are different. And because
$\textbf{\emph{H}}$ is the parity check matrix of perfect code
over $GF(q)$, the number of all $t$th linear combinations of the
$\textbf{\emph{H}}$'s columns satisfies that
\begin{equation}\label{eqn-4}
1 + (q - 1){n \choose 1} + (q - 1)^2 {n \choose 2}   + \cdots + (q
- 1)^t {n \choose t} = q^{n-k}\enspace .
\end{equation}
That means that the set consisting of vectors corresponding to $n$
columns of $\textbf{\emph{H}}$ is a set of $t$th generators of
$GF^{n-k}(q)$. And by Theorem \ref{thm-4}, $\textbf{\emph{H}}$ is
an $(n,n-k,t)$ stego-coding matrix. Furthermore, (\ref{eqn-4})
implies that $\textbf{\emph{H}}$  is a stego-coding matrix of an
MLE code over $GF(q)$.

Conversely, assume $\textbf{\emph{H}}$ is a $(n,n-k,t)$
stego-coding matrix of an MLE code over $GF(q)$. As mentioned in
Subsect. II(A) the rank of $\textbf{\emph{H}}$ is $n-k$, which
implies $\textbf{\emph{H}}$ is a parity check matrix of an $(n,k)$
linear error-correcting code. And by Theorem \ref{thm-4} the set
of vectors corresponding to $n$ columns of $\textbf{\emph{H}}$ is
a set of $t$th generators of $GF^{n-k}(q)$, which, with the fact
that (\ref{eqn-4}) holds by Definition \ref{def-6}, implies that
any two $t$th linear combinations of the $n$ columns of
$\textbf{\emph{H}}$ are different. Therefore the linear code with
$\textbf{\emph{H}}$ as parity check matrix can correct $t$ errors.
Once more by the fact that (\ref{eqn-4}) holds,
$\textbf{\emph{H}}$ is the parity check matrix of a perfect code
over $GF(q)$.
\end{proof}
\end{theorem}
\medskip

\begin{example}\label{exa-2}
Hamming codes are linear single-error-correcting codes. With the
easy decoding method for Hamming codes, we can get easy encoding
method for corresponding stego-codes. For instance, when $q=2$ and
$k=3$, the parity check matrix of binary (7,4) Hamming code is
\[
\textbf{\emph{H}}=\left[\begin{array}{ccccccc} 0 & 0 & 0 & 1 & 1 & 1 & 1 \\ 0 & 1 & 1& 0 & 0 & 1 & 1 \\
1 & 0 & 1 & 0 & 1 & 0 & 1 \end{array} \right] \enspace ,
\]
which is just a (7,3,1) stego-coding matrix and can hides 3 bits
message in a codeword of length of 7 bits by changing at most 1
bit. Here we have taken the columns in the natural order of
increasing binary numbers. For instance, when the inputs are
codeword $\textbf{\emph{x}}=(1,0,0,1,0,0,0)$ and message
$\textbf{\emph{y}}=(1,1,0)$, compute
\[
\textbf{\emph{H}}\textbf{\emph{x}}^{\rm{tr}}=\left[\begin{array}{c}
1\\0\\1
\end{array} \right], \qquad
\left[\begin{array}{c} 1\\0\\1 \end{array} \right] \oplus
\left[\begin{array}{c} 1\\1\\0 \end{array} \right]=
\left[\begin{array}{c} 0\\1\\1 \end{array} \right] \enspace .
\]
Note that the result is the binary representation of 3 and also is
just the third column of $\textbf{\emph{H}}$. Then change the
third position of $\textbf{\emph{x}}$ to output
$\textbf{\emph{x}}'=(1,0,1,1,0,0,0)$ that satisfies
\[
\textbf{\emph{H}}\textbf{\emph{x}}'^{\rm{tr}}=\left[\begin{array}{c}
1\\1\\0
\end{array} \right]= \textbf{\emph{y}}^{\rm{tr}} \enspace .
\]
\end{example}
\bigskip

In fact we can obtain another bound on the dimension of vector
space $GF^k(q)$  by Algorithm \ref{alg-1}.

\medskip
\begin{theorem}\label{thm-10}
If the $t$th dimension of vector space $GF^k(q)$ over $GF(q)$ is
$n$, then
\begin{equation}\label{eqn-5}
n \le \frac{{(q^ {\lfloor {{\scriptstyle{k\over t}}}\rfloor } -
1)(t - 1) + q^{k - \lfloor {{\scriptstyle{k \over t}}} \rfloor (t
- 1)} - 1}}{{q - 1}} \enspace .
\end{equation}
\end{theorem}
\medskip

Because (\ref{eqn-5}) is an upper bound on the $t$th dimension of
vector space $GF^k(q)$, Theorem \ref{thm-5} implies that for any
positive integer $n$ such that
\[n \ge \frac{{(q^ {\lfloor
{{\scriptstyle{k\over t}}}\rfloor } - 1)(t - 1) + q^{k - \lfloor
{{\scriptstyle{k \over t}}} \rfloor (t - 1)} - 1}}{{q - 1}},\]
$(n,k,t)$ linear stego-codes over $GF(q)$ exist.

\section{Nonlinear Stego-codes}
\subsection{Definitions}

The Definition \ref{def-2} for stego-codes is based on
stego-coding function. In fact we can define stego-codes directly
as follows, which is useful for us to study nonlinear stego-codes.

The Hamming distance between two vectors $\textbf{\emph{x}}
\,\mbox{ and }\,\textbf{\emph{y}} \subseteq GF^n(q)$ is denoted by
$\mbox{Dist}(\textbf{\emph{x}} ,\textbf{\emph{y}} )$.

\medskip
\begin{definition}\label{def-7}
By an $M$-partition of $GF^n(q)$, we mean a set $\{I_0 ,I_1 ,
\cdots I_{M-1} \}$ satisfying the following two conditions:
\begin{enumerate}
\item $I_0 ,I_1 , \cdots I_{M-1} $ are non-empty subsets of
$GF^n(q)$ and any two of the $M$ subsets are disjoint; \item
$GF^n(q) = I_0 \cup I_1 \cup \cdots \cup I_{M-1} .$
\end{enumerate}
\end{definition}

\medskip
\begin{definition}\label{def-8}
If $I$ is a nonempty subset of $GF^n(q)$ and $\textbf{\emph{x}}
\in GF^n(q)$, define the distance between $\textbf{\emph{x}} $ and
$I $ as $\mbox{Dist}(\textbf{\emph{x}} ,I) = \mathop {\min
}\limits_{\textbf{\emph{y}} \in I} \mbox{Dist}(\textbf{\emph{x}}
,\textbf{\emph{y}} )$.
\end{definition}

\medskip
\begin{definition}\label{def-9}
An $(n,M,t)$ stego-code over $GF(q)$ is a set ${\cal S} = \{I_0
,I_1 , \cdots ,I_{M-1} \}$ satisfying the following two
conditions:
\begin{enumerate}
\item $\{I_0 ,I_1 , \cdots I_{M-1} \}$ is an $M$-partition of
$GF^n(q)$. \item for any $\textbf{\emph{x}} \in GF^n(q) $ and any
$i$ such that $0 \le i  \le M-1$, $\mbox{Dist}(\textbf{\emph{x}}
,I_i ) \le t$.
\end{enumerate}
\end{definition}
\medskip

For an $(n,M,t)$ stego-code ${\cal S} = \{I_0 ,I_1 , \cdots
I_{M-1} \}$ over $GF(q)$, a corresponding stego-coding function
can be constructed as follows. Let $m = \left\lceil {\log _q M}
\right\rceil $, and the $M$ message symbols can be expressed by
$M$ vectors in $GF^m(q)$, for example, $\textbf{\emph{y}} _0 ,
\cdots ,\textbf{\emph{y}} _{M-1} $. Define function $H:GF^n(q) \to
GF^m(q)$ such that, $H(\textbf{\emph{x}} ) = \textbf{\emph{y}} _i
$, if $\textbf{\emph{x}} \in I_i $, where $0 \le i \le M-1$. Then
with $H$ as decoding function, Definition \ref{def-9} implies that
for any given message $\textbf{\emph{y}} \in GF^m(q)$ and codeword
$\textbf{\emph{x}} \in GF^n(q)$, $\textbf{\emph{y}} $ can be
hidden into $\textbf{\emph{x}} $ (i.e. expressed by
H(\textbf{\emph{x}} )) by changing at most $t$ elements of
$\textbf{\emph{x}} $. Herein $H$ is a vectorial function. And if
every component function of $H$ is a linear function, we call $H$
a linear stego-coding function and call the corresponding code
${\cal S} = \{I_0 ,I_1 , \cdots I_{M-1} \}$ a linear stego-code.
For the linear stego-coding function $H$, if the rank of its
coefficients matrix is $k$, then $\vert I_0 \vert = \vert I_1
\vert = \cdots = \vert I_{M-1} \vert = q^{n - k}$, which means
that $M = q^k$. Therefore the linear stego-code can be simply
denoted by $(n,k,t)$ as we use in Sect. II.

We say that two $(n,M,t)$  stego-codes ${\cal S} = \{I_0 \cup I_1
\cup \cdots \cup I_{M-1} \}$ and $ {\cal T} = \{ J_0 \cup J_1 \cup
\cdots  \cup J_{M-1} \}$ over $GF(q)$ are equivalent if there is a
permutation $\pi $ of the $n$ coordinate positions and $n $
permutations $\sigma _1 , \cdots ,\sigma _n $ of $q$ elements such
that for any $i$ ($0 \le i \le M-1)$, there exists $j$ $(0 \le j
\le M-1)$ satisfying $\pi (\sigma _1 (x_1 ), \cdots ,\sigma _n
(x_n )) \in I_i $ if $(x_1 , \cdots ,x_n ) \in J_j $.

The conclusion in Subsection II(C) implies that there is relations
between the stego-codes and error-correcting codes. The general
definition for error-correcting codes including linear and
nonlinear codes is as follows.

\medskip
\begin{definition}\label{def-10}$^{[8]}$
An $(n,M,d)$ error-correcting code over $GF(q)$ is a
set of $M$ vectors of $GF^n(q)$ such that any two vectors differ
in at least $d$ places, and $d$ is the smallest number with this
property.
\end{definition}
\medskip

To understand the relations and differences between the
error-correcting codes and stego-codes, we think of these codes
geometrically as MacWilliams did in \cite{mac:slo}. The vector
$(a_1 ,a_2 , \cdots ,a_n )$ of length $n$ gives the coordinates of
a vertex of a unit cube in $n$ dimensions. Then An $(n,M,d)$
error-correcting code is just a subset of these vertices while an
$(n,M,t)$ stego-code is a partition of these vertices.

In this geometrical language, the error-correcting coding theory
problem is to choose as many as vertices of the cube as possible
while keeping them a certain distance $d$ apart. However, the
stego-coding theory problem is to divide vertices of the cube as
many disjoint non-empty subsets as possible while keeping any
vertex closer to every subset. In fact, an $(n,M,t)$ stego-code
make the sphere of radius $t$ around any vertex intersects all
these $M$ subsets.

\subsection{Maximum Length Embeddable (MLE) Codes}

With Definition \ref{def-9} of stego-codes, we can generalize
Theorem \ref{thm-6} and Definition \ref{def-6} as following
Theorem \ref{thm-11} and Definition \ref{def-11}.

\medskip
\begin{theorem}\label{thm-11}
An $(n,M,t)$ stego-code over $GF(q)$ must satisfy
\begin{equation}\label{eqn-6}
M \le 1 + (q - 1){n \choose 1} + (q - 1)^2{n \choose 2} + \cdots +
(q - 1)^t{n \choose t} .
\end{equation}
\begin{proof}
Let ${\cal S} = \{I_0 \cup I_1 \cup \cdots \cup I_{M - 1} \}$ be
an $(n,M,t)$ stego-code over $GF(q)$. Then for any given
$\textbf{\emph{x}} \in GF^n(q)$, the sphere of radius $t$ around
$\textbf{\emph{x}} $ must intersect every $I_i $ ($0 \le i \le M -
1)$. Note that this sphere contains $1 + (q - 1){n \choose 1} + (q
- 1)^2{n \choose 2} + \cdots + (q - 1)^t{n \choose t}$ vectors and
these $M$ subsets $I_i \mbox{'s}$ are disjoint, and then we get
the inequality (\ref{eqn-6}).
\end{proof}
\end{theorem}
\medskip

\begin{definition}\label{def-11}
$(n,M,t)$  stego-code over $GF(q)$ is called maximum length
embeddable (abbreviated MLE) if equality holds in (\ref{eqn-6}).
\end{definition}
\medskip

MLE codes have following two interesting properties, and the first
can be obtained from definitions of stego-codes and MLE codes
directly.

\medskip
\begin{lemma}\label{lem-12}
If ${\cal S} = \{I_0 \cup I_1 \cup \cdots \cup I_{M - 1} \}$ is an
MLE $(n,M,t)$ stego-code over $GF(q)$, then for any
$\textbf{\emph{x}} \in GF^n(q)$, the sphere of radius $t$ around
$\textbf{\emph{x}} $ shares only one vector with every $I_i $ ($0
\le i \le M - 1)$.
\end{lemma}
\medskip

\begin{lemma}\label{lem-13}
For the MLE $(n,M,t)$ codes over $GF(q)$, there exists some
integer $k$ such that $M = q^k$.

\begin{proof}
Let ${\cal S} = \{I_0 \cup I_1 \cup \cdots \cup I_{M - 1} \}$ be a
$(n,M,t)$ MLE stego-code over $GF(q)$. Then for any subset $I_i $
($0 \le i \le M - 1)$ and any $\textbf{\emph{x}} \in I_i $, Lemma
\ref{lem-12} implies that, in any $I_j $ ($0 \le j \le M - 1,j \ne
i)$, there is only one vector, for example denote it by
$\textbf{\emph{y}}$, satisfying $\mbox{Dist}(\textbf{\emph{x}}
,\textbf{\emph{y}} ) \le t$. Therefore the mapping $f:I_i \to I_j
$ such that $f(\textbf{\emph{x}} ) = \textbf{\emph{y}} $ if
$\mbox{Dist}(\textbf{\emph{x}} ,\textbf{\emph{y}} ) \le t$ is a
1-1 correspondence between $I_i $ and $I_j $. So there exists
integer $A $ such that $\vert I_0 \vert = \cdots = \vert I_{M-1}
\vert = A$. Assume that the character of field $GF(q)$ is $p $ and
$q = p^r$ , then
\[
AM = A\sum\limits_{i = 0}^t {{n \choose i} (q - 1)^i = q^n
=p^{nr}} .
\]
Therefore there exists some integer $j$ such that $A = p^j$, and
\[
M=\sum\limits_{i = 0}^t {{n \choose i} (q - 1)^i = p^{nr - j}} .
\]
Thus $q - 1 = p^r - 1$ divides $p^{nr - j} - 1$, which implies
that $r$ divides $j$ and $M$ is a power of $q$.
\end{proof}
\end{lemma}
\medskip

In Subsection II(C) we have proved that there is a 1-1
correspondence between linear MLE codes and linear perfect
error-correcting codes. Therefore we guess that there are also
corresponding relations between nonlinear MLE codes and nonlinear
perfect codes.

Hamming bound for error-correcting codes (Lemma \ref{lem-8}) and
the definition of perfect codes has general forms as follows. A
$t$-error-correcting code over $GF(q)$ of length $n $ containing
$M$ codewords must satisfy
\begin{equation}\label{eqn-7}
M\left( 1 + (q - 1){n \choose 1} +  \cdots + (q - 1)^t{n \choose
t}\right) \le q^n .
\end{equation}
If equality holds in (\ref{eqn-7}), the $t$-error-correcting code
over $GF(q)$ of length $n $ containing $M$ codewords is called
perfect code. And it can be proved that the number of codewords of
a perfect code $M$ is a power of $q$ \cite{mac:slo}.

The following two theorems show the relations between the MLE
codes and perfect codes. And we provide two constructive proofs
which can be used to construct MLE codes with perfect codes or
construct perfect codes with MLE codes.

\medskip
\begin{theorem}\label{thm-14}
If $\wp $ is a $t$-error-correcting $(0 \le t \le n)$ perfect code
over $GF(q)$ of length $n $ containing $q^{n - k}$ $(0 \le k \le
n)$ codewords, then there exists a $(n,q^k,t)$ MLE code ${\cal S}
= \{I_0 \cup I_1 \cup \cdots \cup I_{q^k - 1} \}$ over $GF(q)$
such that $\wp $ equals some $I_i $ ($0 \le i \le q^k - 1)$.

\begin{proof}
Let $\wp = \{\textbf{\emph{x}} _1 ,\textbf{\emph{x}} _2 , \cdots
,\textbf{\emph{x}} _{q^{n - k}} \}$ be a $t$-error-correcting
perfect code of length $n$ containing $q^{n - k}$ codewords. Then
the minimum distance of $\wp $ must be larger than $2t$ and $q^{n
- k}\left( {1 + (q - 1){n \choose 1} + \cdots + (q - 1)^t{n
\choose t} } \right) = q^n$. Therefore the number of vectors whose
weights are not larger than $t$ satisfies
\begin{equation}\label{eqn-8}
1 + (q - 1){n \choose 1} + (q - 1)^2{n \choose 1} + \cdots + (q -
1)^t{n \choose 1} = q^k.
\end{equation}
Write these vectors by $\textbf{\emph{y}} _0 , \cdots
,\textbf{\emph{y}} _{q^k - 1} $ and assume that $\textbf{\emph{y}}
_0 $ is the zero vector. Denote the sphere of radius $t$ around
$\textbf{\emph{x}} _i $ by $O_t (\textbf{\emph{x}} _i )$, i.e.
$O_t (\textbf{\emph{x}} _i ) = \{\textbf{\emph{x}} _i +
\textbf{\emph{y}} _j ,\;0 \le j \le q^k - 1\}$ $(1 \le i \le
q^{n-k})$. These $q^{n-k}$ spheres are disjoint because $\wp $ is
a $t$-error-correcting code.

Now construct the stego-code ${\cal S} = \{I_0 \cup I_1 \cup
\cdots \cup I_{M - 1} \}$ as follows.
\begin{equation}\label{eqn-9}
I_i = \{\textbf{\emph{y}} _i + \textbf{\emph{x}} _j ,\;1 \le j \le
q^{n - k}\}, 0 \le i \le q^k - 1.
\end{equation}

We claim that $\{I_0 \cup I_1 \cup \cdots \cup I_{q^k - 1} \}$ is
a partition of $GF^n(q)$. In fact, any two of the $q^k$ subsets
are disjoint. Otherwise, if two subsets, e.g. $I_0 $ and $I_1 $,
are intersectant, then there exist $i\neq j$ such that
$\textbf{\emph{y}} _0 + \textbf{\emph{x}} _i = \textbf{\emph{y}}
_1 + \textbf{\emph{x}} _j $, which implies $O_t (\textbf{\emph{x}}
_i ) \cap O_t (\textbf{\emph{x}} _j ) \ne \emptyset $, and a
contradiction to $O_t (\textbf{\emph{x}} _i )$'s being disjoint
follows. Furthermore note that every $I_i$ $(0 \le i \le q^k - 1)$
contains $q^{n - k}$ vectors. Therefore $GF^n(q) = I_0 \cup I_1
\cup \cdots \cup I_{q^k - 1} $.

Now to prove $\{I_0 \cup I_1 \cup \cdots \cup I_{q^k - 1} \}$
being a stego-code, the only thing we should verify is that for
any $\textbf{\emph{z}} \in GF^n(q)$, the sphere of radius $t$
around $\textbf{\emph{z}} $, i.e. $O_t (\textbf{\emph{z}} ) =
\{\textbf{\emph{z}} _j :\textbf{\emph{z}} _j = \textbf{\emph{z}} +
\textbf{\emph{y}} _j \,\mbox{and}\;0 \le j \le q^k - 1\}$,
intersects every $I_i$ $(0 \le i \le q^k - 1)$. Otherwise, there
must exist some subset, e.g. $I_h $, that shares at least two
vectors with $O_t (\textbf{\emph{z}} )$ because $O_t
(\textbf{\emph{z}} )$ includes only $q^k$ vectors. For instance,
if there are $0 \le i_1 < i_2 \le q^k - 1$ such that
$\textbf{\emph{z}} _{i_1 } \in I_h $ and $\textbf{\emph{z}} _{i_2
} \in I_h $, then there exist $0 \le j_1 < j_2 \le q^{n - k}$ such
that $\textbf{\emph{z}} _{i_1 } = \textbf{\emph{y}} _h +
\textbf{\emph{x}} _{j_1 } $ and $\textbf{\emph{z}} _{i_2 } =
\textbf{\emph{y}} _h + \textbf{\emph{x}} _{j_2 } $. Therefore, on
one hand, $\mbox{Dist}(\textbf{\emph{z}} _{i_1 }
,\textbf{\emph{z}} _{i_2 } ) = \mbox{Dist}(\textbf{\emph{z}} +
\textbf{\emph{y}} _{i_1 } ,\textbf{\emph{z}} + \textbf{\emph{y}}
_{i_2 } ) = \mbox{Dist}(\textbf{\emph{y}} _{i_1 }
,\textbf{\emph{y}} _{i_2 } ) \le 2t$, but on the other hand,
$\mbox{Dist}(\textbf{\emph{z}} _{i_1 } ,\textbf{\emph{z}} _{i_2 }
) = \mbox{Dist}(\textbf{\emph{y}} _h + \textbf{\emph{x}} _{j_1 }
,\textbf{\emph{y}} _h + \textbf{\emph{x}} _{j_2 } ) =
\mbox{Dist}(\textbf{\emph{x}} _{j_1 } ,\textbf{\emph{x}} _{j_2 } )
> 2t$. And a contradiction follows. So we prove that $\{I_0 \cup
I_1 \cup \cdots \cup I_{q^k - 1} \}$ is an $(n,q^k,t)$ stego-code,
and it is a MLE code because (\ref{eqn-8}) holds. Finally,
(\ref{eqn-9}) means $I_0 = \wp $, because $\textbf{\emph{y}} _0 $
is the zero vector.
\end{proof}
\end{theorem}
\medskip

\begin{theorem}\label{thm-15}
If ${\cal S} = \{I_0 \cup I_1 \cup \cdots \cup I_{q^k - 1} \}$ is
an $(n,q^k,t)$ MLE code over $GF(q)$, then every $I_i $ ($0 \le i
\le q^k - 1)$ is a $t$-error-correcting perfect code over $GF(q)$
of length $n$ containing $q^{n - k}$ codewords.

\begin{proof}
Let ${\cal S} = \{I_0 \cup I_1 \cup \cdots \cup I_{M - 1} \}$ be
an $(n,q^k,t)$ MLE code over $GF(q)$. The poof of Lemma
\ref{lem-13} implies that every $I_i $ ($0 \le i \le q^k - 1)$
contains $q^{n - k}$ vectors. Now we prove any $I_i $, e.g. $I_0
$, is a $t$-error-correcting code. In fact, for any two vectors
$\textbf{\emph{x}} _1 ,\textbf{\emph{x}} _2 \in I_0 $, the sphere
of radius $t$ around them, i.e. $O_t (\textbf{\emph{x}} _1 )$ and
$O_t (\textbf{\emph{x}} _2 )$, are disjoint. Otherwise, if there
exists $\textbf{\emph{z}} \in O_t (\textbf{\emph{x}} _1 ) \cap O_t
(\textbf{\emph{x}} _2 )$, then the sphere of radius $t$ around
$\textbf{\emph{z}} $ shares two vectors with $I_0 $, which is
contrary to Lemma \ref{lem-12}. Therefore $I_0 $ is a
$t$-error-correcting code of length $n $ containing $q^{n - k}$
codewords. Furthermore, because ${\cal S} = \{I_0 \cup I_1 \cup
\cdots \cup I_{M - 1} \}$ is an MLE code, $q^{n - k}\left( 1 + (q
- 1){n \choose 1}+ \cdots + (q - 1)^t {n \choose t} \right) = q^{n
- k}q^k = q^n$, which implies that $I_0 $ is a perfect code.
\end{proof}
\end{theorem}
\medskip

Theorem \ref{thm-14} and \ref{thm-15} show that there is a
corresponding relation between perfect codes and MLE codes in
equivalent sense. And in fact these two theorems imply that the
classifications of MLE codes can be determined by the
classifications of perfect codes.

There are there kinds of trivial perfect codes: a code containing
just one codeword, or the whole space, or a binary repetition code
of odd length. We call the corresponding MLE codes also trivial
MLE codes, i.e. $(n,q^n,n)$ or ($n$, 1, 0) code over $GF(q)$, or
binary $(2t + 1,2^{2t},t)$ code, which can be constructed by
Theorem \ref{thm-14}.

The work of Tiet$\ddot{a}$v$\ddot{a}$ine \cite{tie} shows that
there are only there kinds of parameters $n$, $M$ and $d$ for
nontrivial perfect codes.
\begin{enumerate}
\item The binary $(23,2^{12},7)$ Golay code (linear
three-error-correcting code) which is unique in the sense of
equivalence. \item The ternary $(11,3^6,5)$ Golay code (linear
two-error-correcting code) which is unique in the sense of
equivalence. \item The $\left( \frac{q^r - 1}{q - 1},q^{\frac{q^r
- 1}{q - 1} - r},3 \right)$ code over $GF(q)$
(single-error-correcting code). All linear perfect codes with
these parameters are equivalent, i.e. the Hamming codes. And there
exist nonlinear perfect codes with these parameters over $GF(q)$
for all $q$.
\end{enumerate}

Correspondingly, Theorem \ref{thm-14} and \ref{thm-15} imply that
there are also only there kinds of possible parameters $n$, $M$
and $t$ for MLE nontrivial codes.

\medskip
\begin{corollary}\label{cor-16}
An MLE codes must belong to one of the following three types:
\begin{enumerate}
\item The binary linear $(23,2^{11},3)$ code. All MLE codes with
these parameters are equivalent. \item The ternary linear
$(11,3^5,2)$ code. All MLE codes with these parameters are
equivalent. \item The $\left( \frac{q^r - 1}{q - 1},q^r,1 \right)$
code over $GF(q)$. All linear MLE codes with these parameters are
equivalent. And there exist nonlinear MLE codes with these
parameters over $GF(q)$ for all $q.$
\end{enumerate}
\end{corollary}
\medskip

For the security of steganographic systems, we hope there are
enough stego-codes, especially binary codes. And the following
corollary shows that there are indeed so many binary MLE codes. In
fact, Krotov \cite{kro} ever proved that there are at least
\[2^{2^{\frac{n +1}{2} - \log _2 (n + 1)}} \cdot 3^{2^{\frac{n - 3}{4}}} \cdot
2^{2^{\frac{n + 5}{4} - \log _2 (n + 1)}}
\]
different perfect binary codes of length $n$ $(n = 2^r - 1)$.
Therefore, with Theorem \ref{thm-14} and \ref{thm-15} we can
obtain the following lower bound for length $n$ binary MLE codes.

\medskip
\begin{corollary}\label{cor-17}
There are at least
\[
\frac{2^{2^{\frac{n + 1}{2} - \log _2 (n + 1)}} \cdot
3^{2^{\frac{n - 3}{4}}} \cdot 2^{2^{\frac{n + 5}{4} - \log _2 (n +
1)}}}{n + 1}
\]
different MLE binary codes of length $ n$, where $n = 2^r - 1$.
\end{corollary}
\medskip

So far there have been many designs for different nonlinear
perfect binary codes with which and Theorem \ref{thm-14} we can
construct the corresponding MLE binary codes.

\section{Hiding Redundancy -- The Performance of Stego-codes}

Usually the performance of encoding method for steganography is
valued by `` message rate'', ``change density'' or ``embedding
efficiency''. For example, for the sequential  LSB steganography
on images, we say that the message rate is 100\% (the LSB of every
pixel carries one bit message), the change density is 50\% (on
average 50\% pixels needn't to be changed), and so the embedding
efficiency is 2 (on average embed 2 bits per change). However
these three measures can only reflect one aspect of this problem
respectively. In fact, the user hopes to get the maximum message
rate within a proper constraint of ``change density'', which is
just the so called hiding capacity. Therefore the difference
between the hiding capacity and message rate, which we call as
``hiding redundancy'' in this paper, can reflects the capability
of a stego-code soundly. To introduce the concept of hiding
redundancy, the following preparations are needed.

We use the following notations. Random variables are denoted by
capital letters (e.g. $X$), and their realizations by respective
lower case letters (e.g. $x$). The domains over that random
variables are defined are denoted by script letters (e.g.
$\mathcal{X}$). Sequences of $N$ random variables are denoted with
a superscript (e.g. $X^{N}=(X_{1},X_{2},\cdots ,X_{N})$ which
takes its values on the product set $\mathcal{X}^{N}$). And we
denote entropy and conditional entropy with $H( \cdot )$ and
$H(\cdot|\cdot)$ respectively.

Assume that the cover-objects $\widetilde{X}^N$ are independent
and identically distributed (i.i.d) samples from
$P(\widetilde{x})$. Because the embedded message $M$ usually is
cipher text, we assume that it is uniformly distributed, and
independent of $\widetilde{X}^N$. And $M$ is hidden in
$\widetilde{X}^N$, in the control of a secret stego-key $K$,
producing the stego-object $X^N$.

A formal definition of steganographic system (abbreviated
stegosystem) is present by Moulin \cite{mou:sul}. First of all,
the embedding algorithm of a stegosystem should keep transparency
that can be guaranteed by some distortion constraint. A distortion
function is a nonnegative function $d:{\cal X} \times {\cal X} \to
{\cal R}^ + \cup \{0\}$, which can be extended to one on
$N$-tuples by $d(x^N ,y^N ) = \frac{1}{N}\sum_{i = 1}^N {d(x_i
,y_i )}$. A length-$N$ stegosystem \footnote{In \cite{mou:sul} the
terms of information hiding code is used here. To distinguish the
problem of this paper and that of \cite{mou:sul}, we replace it by
stegosystem.} subject to distortion $D$ is a triple $({\cal M},f_N
,\phi _N )$, where ${\cal M}$ is the message set, $f_N :{\cal X}^N
\times {\cal M} \times {\cal K} \to {\cal X}^N$ is the embedding
algorithm subject to the distortion constraint $D$, and $\phi _N
:{\cal X}^N  \times {\cal K}  \to {\cal M}$ is the extracting
algorithm.

A cover channel is a conditional $p.m.f.$ (probability mass
function) $q(x|\widetilde{x}):{\cal X} \to {\cal X}$. Denote the
set of cover channels subject to distortion $D$ by $Q$.
Furthermore, define the message rate as $R_m  = \frac{{H(M)}}{N}$
and the probability of error as $P_{eN}=P(\phi _N (X^N ,K) \ne
M)$.

The hiding capacity is the supremum of all achieve message rates
of stegosystems subject to distortion $D$ under the condition of
zero probability of error (i.e. $P_{e,N} \to 0 \mbox{ as } N \to
\infty $). When disregarding the active attacker, the results of
\cite{mou:sul,mou:wan} imply that the expression of hiding
capacity for stegosystem can be given by
\begin{equation}\label{eqn-10}
C(D) = \mathop {\max }\limits_{q(x|\tilde{x}) \in Q}
H(X|\widetilde{X})\enspace .
\end{equation}

Because $C(D)$ is the maximum of the conditional entropy through
all cover channels subject to $D$, $C(D)$ just reflects the hiding
ability of the cover-object within the distortion constraint. So
we refer to $C(D)-R_m$ as the hiding redundancy of cover-objects,
which can reflect the hiding capability of a stegosystem. We have
assumed that the embedded message is uniformly distributed, and
independent of $\widetilde{X}^N$, which means that there are
uniformly distributed values at the positions to be changed. Then
an $(n,k,t)$ stego-coding function and a corresponding encoding
algorithm can compose a stegosystem with message rate being
$\frac{k}{n}$. And when using Hamming distance as distortion
function, the average distortion is just the change density.
However note that $\frac{t}{n}$ is the maximum distortion. And the
computation of average distortion relies on the encoding
algorithm. For the linear $(n,k,t)$ steg-code over $GF(2)$ , as
mentioned in Sect. II, its encoding algorithm can be formulated as
a table consisting of $2^k$ $n$-dimension vectors. Let $a_i$,
where $0 \le i \le t$ , be the number of vectors of weight $i$ in
the table. Then the average distortion (change density) of this
code is $\frac{1}{2^k }\sum_{i = 1}^t {a_i \frac{i}{n}} $. For
instance, the average distortion of $(2^k -1,k,1)$ stego-code in
F5 (Example \ref{exa-1}) equals $\frac{1}{2^k }[1 \cdot
\frac{0}{2^k - 1}+ (2^k -1)\cdot \frac{1}{2^k  - 1}] =
\frac{1}{2^k }$.

It is hard to compute the hiding capacity for general
cover-objects. Now consider Bernoulli$(\frac{1}{2})$-Hamming case:
The set of symbols of cover-objects is $\mathcal{X}=\{0,1\}$ , and
the sequence of cover-objects $\widetilde{X}^N$ satisfies
distribution of Bernoulli$(\frac{1}{2})$; The distortion function
is Hamming distance, i.e. $d(x,y) = x \oplus y$. The hiding
capacity for this case has been given by \cite{mou:wan}.

\medskip
\begin{lemma}\label{lem-19}
$^{[12]}$ For Bernoulli$(\frac{1}{2})$-Hamming case with
distortion constraint $D$, the hiding capacity is
\[
C(D)=\left\{\begin{array}{ll} H(D) \quad & \mbox{if  } 0 \le D \le
\frac{1}{2} \\ 1 \quad & \mbox{if  } D > \frac{1}{2} \end{array}
\right . \enspace ,
\]
where $H(D)= -D\log_2 D - (1 - D)\log _2 (1 - D)$.
\end{lemma}
\medskip

LSBs of images satisfies distribution of Bernoulli$(\frac{1}{2})$
approximatively. So we take LSB steganography as a criterion, i.e.
apply stego-codes to LSB steganography, to compare the performance
of different stego-codes.

\medskip
\begin{example}[\rmfamily {Hiding Redundancy of Stego-codes}]\label{exa-3}
For the simple LSB steganography, the message rate is $2D$ when
distortion is $D$ and $0 \le D \le \frac{1}{2}$, therefore the
hiding redundancy is $H(D)-2D$. On the other hand, for the $(2^k -
1,k,1)$ stego-code in F5, the message rate is $\frac{k}{2^k-1}$,
distortion is $\frac{1}{2^k}$, and then the hiding redundancy is
$H(\frac{1}{2^k})-\frac{k}{2^k-1}$. Fig. \ref{fig:redundancy2}
shows that F5 is better than simple LSB steganography, because the
hiding redundancy of F5 is smaller.

\begin{figure}
\centering \includegraphics[width=2.5in]{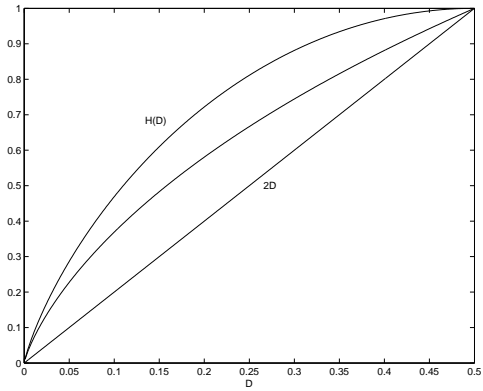}
\caption{The middle curve is obtained by connecting the points
such as $(\frac{1}{2^k}, \frac{k}{2^k-1})$. The difference between
the two curves is the hiding redundancy of F5; and the difference
between the curve of $H(D)$ and the beeline of $2D$ is the hiding
redundancy of simple LSB steganography.} \label{fig:redundancy2}
\end{figure}
\end{example}
\medskip

Furthermore, by Lemma \ref{lem-19}, we can get another bound on
the length of binary stego-codes.

\medskip
\begin{theorem}\label{thm-20}
The $(n,k,t)$ steg-code over $GF(2)$ such that $\frac{t}{n} \le
\frac{1}{2}$ must satisfy
\[
\frac{k}{n} \le H\left(\frac{t}{n}\right) \enspace .
\]
\begin{proof}
For any given $(n,k,t)$ steg-code over $GF(2)$, assume its average
distortion (change density) is $D$. By the definition of capacity,
the message rate $\frac{k}{n}$ is smaller than the hiding capacity
$C(D)$. And when $\frac{t}{n} \le \frac{1}{2}$, we have $H(D) \le
H(\frac{t}{n})$ because $D \le \frac{t}{n}$ (Note that
$\frac{t}{n}$ is the maximum distortion). Apply this code to the
cover-object satisfying distribution of Bernoulli$(\frac{1}{2})$
and Lemma \ref{lem-19} implies that $\frac{k}{n} \le C(D) = H(D)
\le H\left( {\frac{t}{n}} \right)$.

\end{proof}
\end{theorem}
\medskip

Specially for linear binary stego-codes, combining Theorem
\ref{thm-5} and \ref{thm-20}, we can get the following interesting
result which seems hard to be obtained from the point of view of
algebra directly.

\medskip
\begin{corollary}\label{cor-21}
If the $t$th $(1 \le t \le k)$ dimension of vector space $GF^k(2)$
over $GF(2)$ is $n$ and $\frac{t}{n} \le \frac{1}{2}$, then
\[
\frac{k}{n} \le H\left(\frac{t}{n}\right) \enspace .
\]
\end{corollary}

\section{Conclusions}
In this paper, we formally define the stego-code that is a new
coding problem, and studied the construction and properties of
this kind of code. However there are still many interesting
problems about this topic, such as the estimation of $t$th
dimension and the construction of minimum set of $t$th generators
of $GF^k(q)$, other bounds on the length of stego-codes, the
construction of fast encoding algorithms, the construction of
codes that can approach the hiding capacity, and the further
relations between stego-codes and error-correcting codes. Further
researches also include the applications of stego-codes in other
possible fields.

\appendices



\section*{Acknowledgment}
This paper is supported by NSF of China No. 60473022. And the
authors would like to thank Wenfen Liu and Jia Cao for many
helpful and interesting discussions.



%

\begin{biography}{Weiming Zhang}
was born in Hebei, P. R. China in 1976. He is working for the Ph.D
degree in Cryptology in Zhengzhou Information Engineering
University. His research interests include probability theory,
cryptology, and information hiding.
\end{biography}

\begin{biography}{Shiqu Li}
 was born in Sichuan, P. R. China in 1945. He received his
MSc. degrees in probability theory from Beijing Normal University,
P. R. China in 1981. He is currently a Professor in the Department
of Applied Mathematics at Zhengzhou Information Engineering
University. His primary research interests include probability
theory, cryptology, and especially the logic funtions in
cryptology.
\end{biography}



\vfill


\end{document}